\documentclass[aps,prl,reprint,superscriptaddress]{revtex4-1}

\usepackage{graphicx}
\usepackage{amsmath}
\usepackage{textcomp}
\usepackage{xcolor}
\usepackage{ulem}

\begin{document}

\title{Dynamic response of an artificial square spin ice}

\author{M.~B.~Jungfleisch}
\email{jungfleisch@anl.gov}
\affiliation{Materials Science Division, Argonne National Laboratory, Argonne IL 60439, USA}

\author{W.~Zhang}
\affiliation{Materials Science Division, Argonne National Laboratory, Argonne IL 60439, USA}

\author{E.~Iacocca}
\affiliation{Department of Applied Mathematics, University of Colorado at Boulder, Boulder CO 80309, USA}
\affiliation{Department of Applied Physics, Division for Condensed Matter Theory, Chalmers University of Technology, 412 96 Gothenburg, Sweden}

\author{J.~Sklenar}
\affiliation{Materials Science Division, Argonne National Laboratory, Argonne IL 60439, USA}
\affiliation{Department of Physics and Astronomy, Northwestern University, Evanston IL 60208, USA}

\author{J.~Ding}
\affiliation{Materials Science Division, Argonne National Laboratory, Argonne IL 60439, USA}

\author{W.~Jiang}
\affiliation{Materials Science Division, Argonne National Laboratory, Argonne IL 60439, USA}

\author{{S.~Zhang}}
\affiliation{Materials Science Division, Argonne National Laboratory, Argonne IL 60439, USA}
 
\author{J.~E.~Pearson}
\affiliation{Materials Science Division, Argonne National Laboratory, Argonne IL 60439, USA}

 \author{V.~Novosad}
 \affiliation{Materials Science Division, Argonne National Laboratory, Argonne IL 60439, USA}
 
\author{J.~B.~Ketterson}
\affiliation{Department of Physics and Astronomy, Northwestern University, Evanston IL 60208, USA}

 \author{O.~Heinonen}
\affiliation{Materials Science Division, Argonne National Laboratory, Argonne IL 60439, USA}
\affiliation{Northwestern-Argonne Institute of Science and Engineering, Northwestern University, IL 60208, USA}

\author{A.~Hoffmann}
\affiliation{Materials Science Division, Argonne National Laboratory, Argonne IL 60439, USA}

\date{\today}

\begin{abstract}

Magnetization dynamics in an artificial square spin-ice lattice made of Ni$_{80}$Fe$_{20}$ with magnetic field applied in the lattice plane is investigated by broadband ferromagnetic resonance spectroscopy. The experimentally observed dispersion shows a rich spectrum of modes corresponding to different magnetization states. These magnetization states are determined by exchange and dipolar interaction between individual islands, as is confirmed by a semianalytical model. In the low field regime below 400 Oe a hysteretic behavior in the mode spectrum is found. Micromagnetic simulations reveal that the origin of the observed spectra is due to the initialization of different magnetization states of individual nanomagnets. Our results indicate that it might be possible to determine the spin-ice state by resonance experiments and are a first step towards the understanding of artificial geometrically frustrated magnetic systems in the high-frequency regime.

\end{abstract}
\maketitle

\begin{figure}[b]
\includegraphics[width=1\columnwidth]{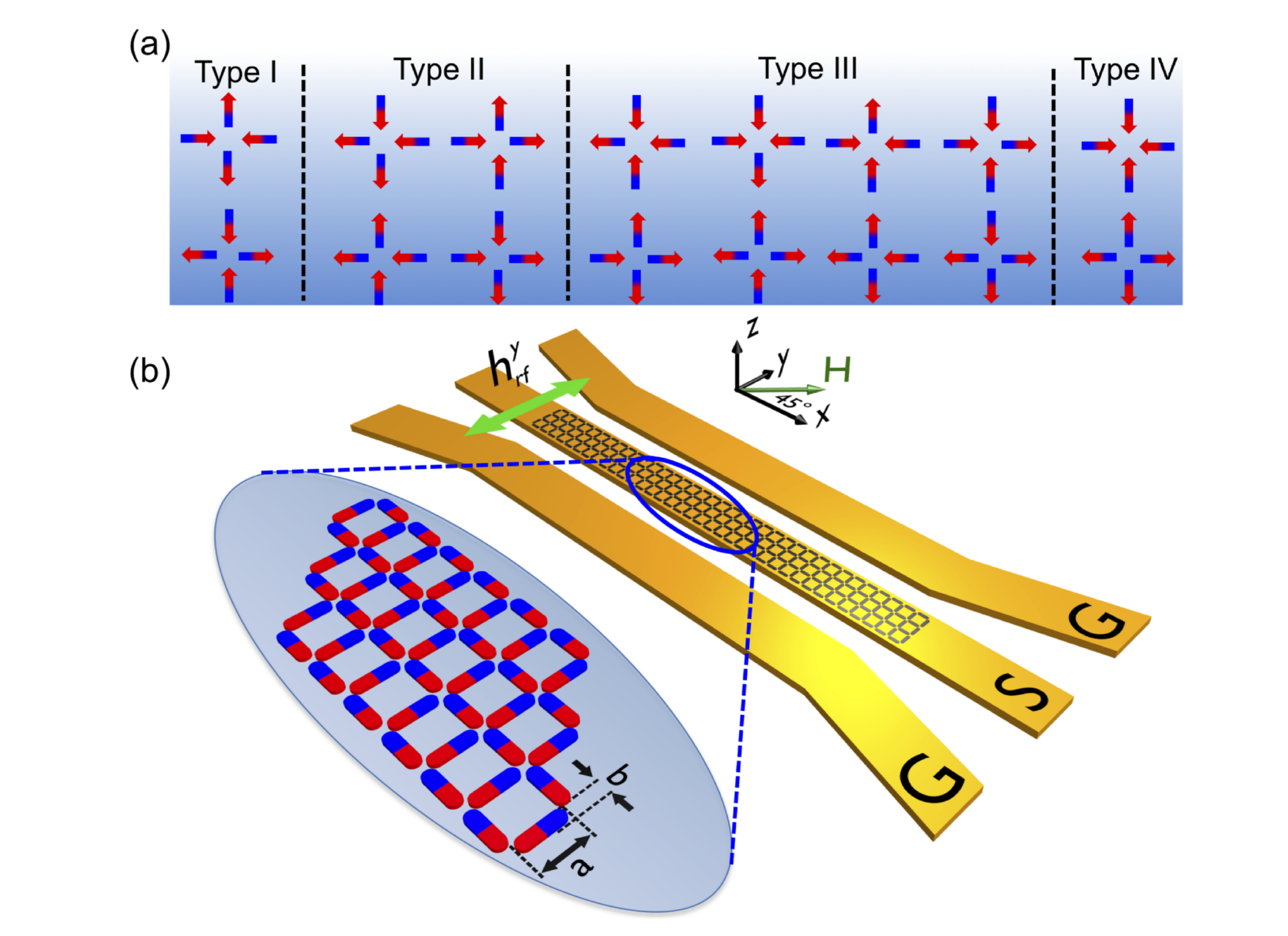}
\caption{\label{Fig1} (Color online) (a) Illustration of the $16$ possible moment configurations of a square spin-ice lattice. (b) Illustration of the experimental setup and measurement configuration. The artificial spin ice is patterned on the signal {line}. The bias magnetic field is oriented along the long side of the CPW ($x$-direction). The inset shows a sketch of the artificial square spin-ice lattice. The islands have the dimensions: $a=300$~nm and $b=130$~nm, thickness $20$~nm.} 
\end{figure}

Frustrated magnetic systems, such as spin ices, have been of scientific interest for a long time due to their highly degenerated ground states, which result in complex magnetic ordering and collective behavior \cite{Nisoli, Schiffer_Nature,Cumings,Braun,Morgan}. In contrast to the prototypical crystalline materials that started the exploration of spin-ice systems, such as the pyrochlores Dy$_2$Ti$_2$O$_7$
, Ho$_2$Ti$_2$O$_7$ 
 and Ho$_2$Sn$_2$O$_7$ 
 \cite{Harris,pyrochlore}, artificially structured spin-ice lattices offer the unique opportunity to control and engineer the interactions between the elements by their geometric properties and orientation \cite{Nisoli,Wang_Nature_2006,Heydermann_2013}. Another outstanding advantage of artificial spin ices is that the magnetization state of each individual \textit{spin} (i.e., macrospin/single domain magnetic particle) is directly accessible through magnetic microscopy (e.g., scanning probe, electron, optical or X-ray microscopy). The 16 possible magnetization configurations of a square spin ice are shown in Fig.~\ref{Fig1}(a).

Spin dynamics in magnonic crystals, materials with periodic perturbations or variations in one of the magnetic properties of the system, have been extensively investigated \cite{Krawczyk,Chumak,Ding,Ding_APL}. One- and two-dimensional magnonic crystals were studied and the research community paid particular attention to nano-structured materials \cite{Krawczyk}, such as chains of dots or arrays of discs \cite{Huber}, antidot lattices with different shapes and alignments \cite{Neusser_PRL,Neusser_PRB11,Jungfleisch_APL_2015}, gratings or nanostripes \cite{Adeyeye}, etc.

Although artificial spin ices offer a fascinating playground to investigate how specific magnetization states of individual islands or defects can affect the collective spin dynamics, there are only very few works on dynamics in the GHz-regime \cite{Joe_JAP13,Olle_PRL} reported. Sklenar \textit{et al}. show broadband ferromagnetic resonance (FMR) measurements on an artificial bicomponent square spin-ice lattice utilizing a macroscopic meanderline approach and find a field-dependent behavior in remanence where interactions between individual elements presumably play a less important role. 
Furthermore, the geometrical arrangement of the structures in the artificial lattice leads to frustration by design and it was shown by micromagnetic simulations that Dirac strings between topological defects in these systems are expected to exhibit complex magnetization dynamics and characteristic signatures in the mode spectrum \cite{Olle_PRL}.

In this Letter, we present results on magnetization dynamics of an artificial square spin ice characterized by broadband FMR spectroscopy. These observations are confirmed by a semianalytical model where each element is divided in three exchange coupled ``macrospins''. Furthermore, a hysteretic behavior of modes in the low bias magnetic field regime is found, where the system is not in remanence. In order to interpret our results, we link our findings to micromagnetic simulations and find that the hysteresis in the mode spectrum relates to the magnetization configuration of particular islands that evolves from specific applied field routines. 
Comparing the experimental results with simulations, we find evidence to determine the spin-ice state [Fig.~\ref{Fig1}(a)] from the resonance spectra. Additionally, the simulations corroborate the experimentally found spectra and the results of the semianalytical model. 

\begin{figure}[t]
\includegraphics[width=1\columnwidth]{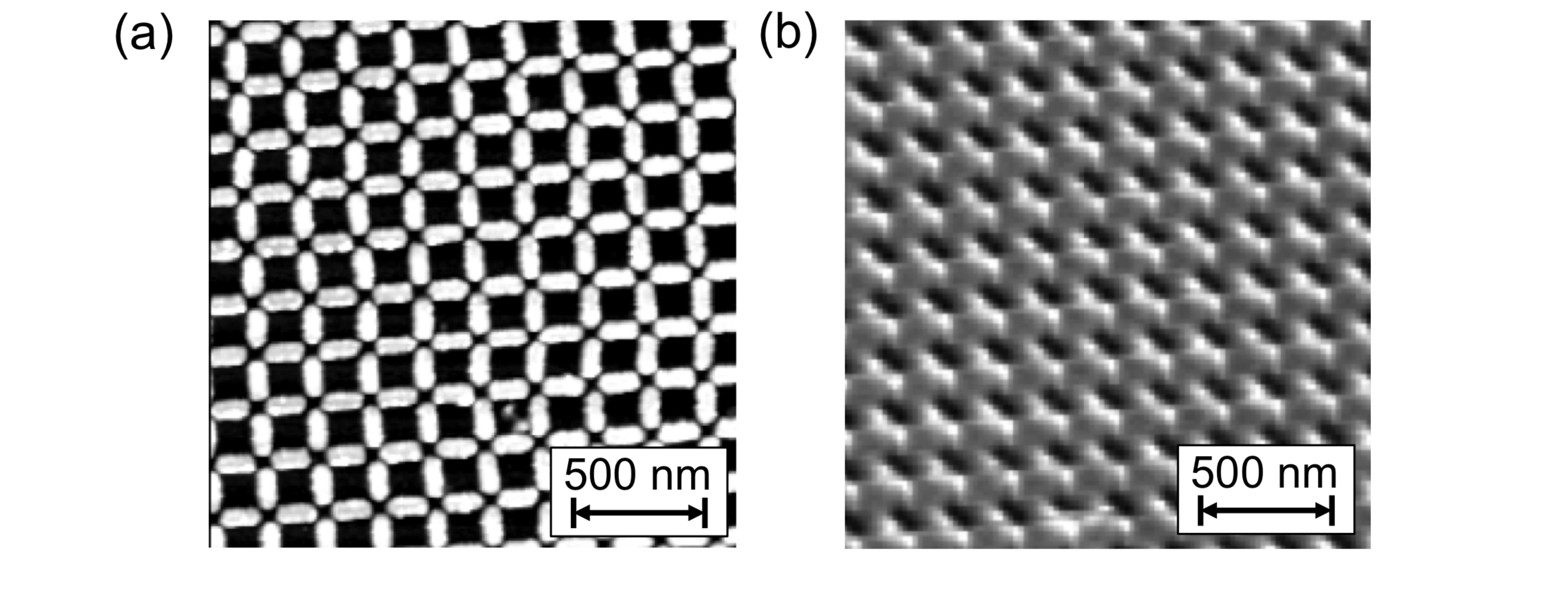}
\caption{\label{Fig2} (Color online) (a) AFM and (b) corresponding MFM image. The sample was saturated at $-3100$~Oe before the MFM characterization.} 
\end{figure}
%

Figure~\ref{Fig1}(b) illustrates a sketch of the experimental setup. The spin-ice lattice was patterned by electron beam lithography on a  coplanar waveguide (CPW). The external magnetic field is aligned in the $x$-direction. Figure~\ref{Fig2} shows the actual spin-ice lattice made of 20~nm thick permalloy (Py, Ni$_{80}$Fe$_{20}$). Details about the sample fabrication can be found in the supplementary materials (SM). In Fig.~\ref{Fig2}(a) the topography of the sample measured by atomic force microscopy (AFM) is depicted and Fig.~\ref{Fig2}(b) shows the corresponding magnetic force microscopy (MFM) image. 
Individual islands have lateral dimensions of  $\sim 300\times 130$~nm$^2$ and are separated by a gap of approximately 30~nm, see SM. 
\begin{figure}[t]
\includegraphics[width=1\columnwidth]{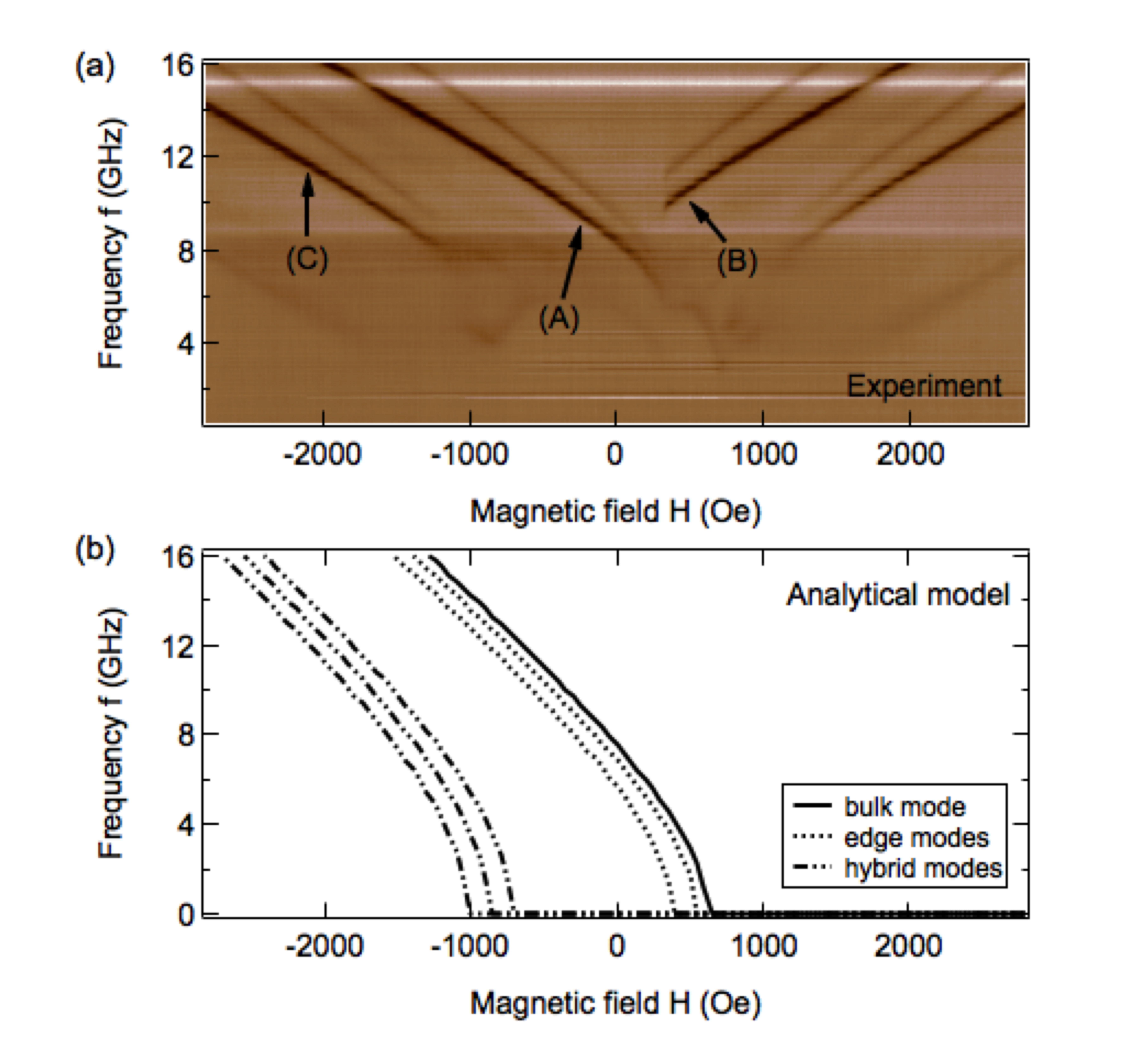}
\caption{\label{Fig3} (Color online) (a) Ferromagnetic resonance measurement on artificial square spin ice, shown in Fig.~\ref{Fig2}. The background signal is taken at $-3100$~Oe and subtracted from the spectrum. The magnetic field is swept from negative to the positive field direction. (b) Field-dependent modes obtained from the semianalytical model. The islands are artificially magnetized parallel to a negative field.} 
\end{figure}
 
\begin{figure*}[t]
\includegraphics[width=2\columnwidth]{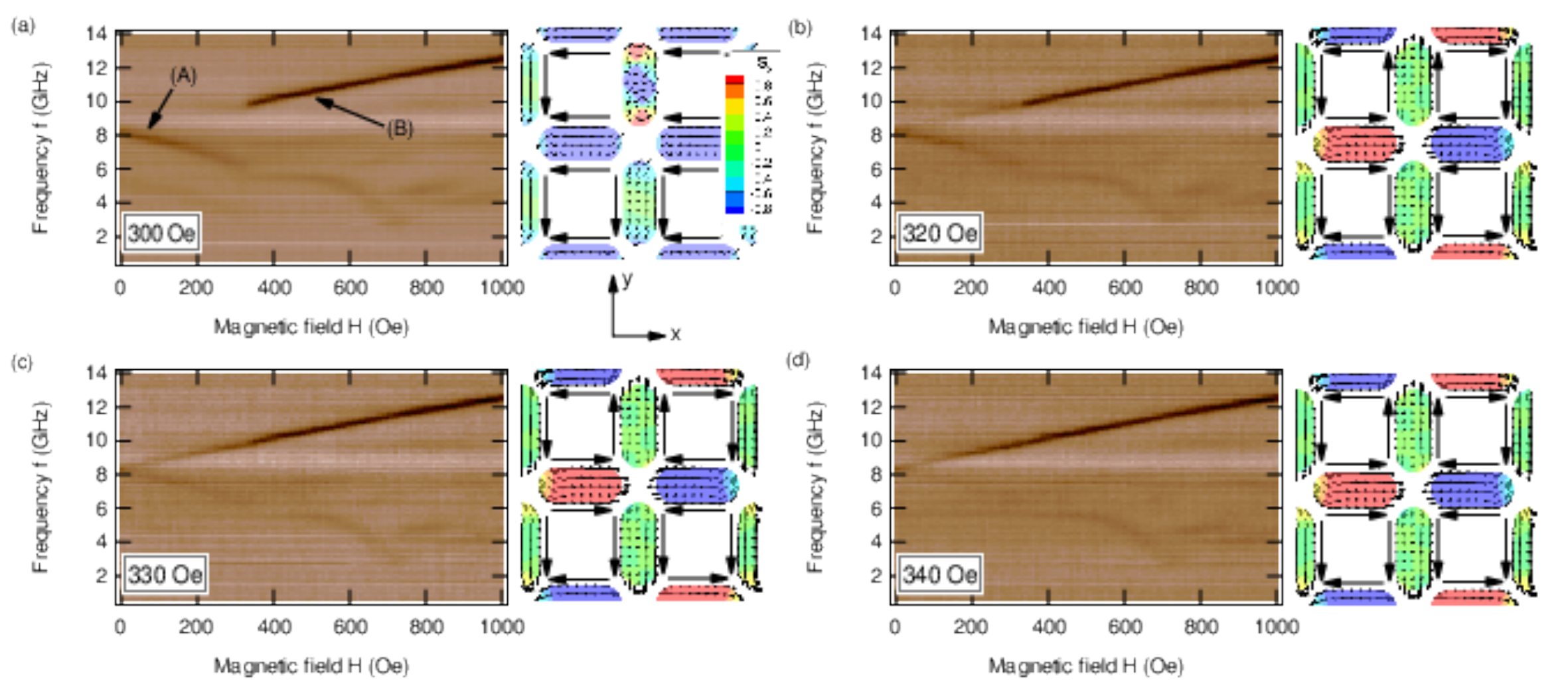}
\caption{\label{Fig4} (Color online) Observation of FMR spectra in the low magnetic field regime. After saturating at $-3100$~Oe, the initialization field $H_\mathrm{ini}$ is applied and then the spectrum is recorded starting at 0 Oe while gradually increasing the applied field. (a)-(d) $H_\mathrm{ini}$ was varied from 300 Oe to 340 Oe. The static equilibrium magnetization states at zero-field are shown next to each spectrum. The color coding indicates the horizontal component of the magnetization going from blue (left or negative x-direction) to red (right or positive x-direction). The arrows next to each island indicate the average magnetization direction in the island.} 
\end{figure*}
 The measurement configuration is shown in Fig.~\ref{Fig1}(b). The microwave driven Oersted field is aligned perpendicularly to the external magnetic field 
and the spin-ice lattice is oriented parallel to the signal line, as shown in Fig.~\ref{Fig1}. 
We employ a broadband vector network analyzer FMR (VNA-FMR) technique \cite{Kalarickal} to measure the dynamics response. 
 For each magnetic field step, an averaged VNA spectrum is recorded. 
The spectrum is shown in a false color-coded image in Fig.~\ref{Fig3}(a). Here, the magnetic field was swept from negative to positive magnetic fields. From the spectrum we can draw several conclusions: First of all, we observe a rich mode spectrum. 
Secondly, at high frequencies above 6 GHz multiple modes are detected, which will be identified below as Kittel-like modes using a semianalytical model and micromagnetic simulations. 
Thirdly, an asymmetry of the spectrum around zero-field is found. If the field is swept from positive to negative fields rather than from negative to positive fields, the spectrum is mirrored around zero-field (not shown). 
 
Further insight can be gained from a semianalytical model of our structure. Following the method outlined in Ref.~[\onlinecite{Iacocca_PRB}], it is possible to find the eigenvalues excited in a magnetized square ice taking into account the effect of external fields, shape anisotropy, and dipolar coupling. This is accomplished by reducing the 
structure to a unit cell composed of two magnetic elements {in case of the remanent state with nonzero magnetization (apply field along (1,1) direction, then reduce field to zero)} where Bloch waves are excited. {Contrarily, the spin ice ground state (type I) has four magnetic elements per unit cell.} In order to better represent the dynamics, the model assumes that each magnetic element in the structure is divided in three exchange-coupled ``macrospins'' with an effective coupling strength proportional to their volume and the exchange stiffness of Py, $A = 1~ \mu$erg/cm. 
Consequently, we introduce {six} degrees of freedom and hence six possible eigenvalues. Setting a magnetized state, {i.e.}, all spins pointing along the direction of a {negative} field, we obtain the field-dependent frequencies of the coherent spin-wave mode ($\vec{k}=0$) shown Fig.~\ref{Fig3}(b). We find an excellent qualitative agreement to experiments using this simplified model without any fitting parameters. In particular, the semianalytical solution correctly reproduces the spectrum hysteresis, the qualitative form of the field-dependent frequencies, and the existence of two bands split at non-harmonic frequencies. From {this approach}, it is possible to spatially locate the modes. We find that the higher frequency branch {(A)} corresponds to the mode dominated by magnetic elements whose easy axis lies parallel to the applied field whereas the lower band {(C)} corresponds to elements magnetized perpendicular to its easy axis. This agrees with the fact that the anisotropy field in the hard axis opposes the applied field and therefore reduces the frequency. Furthermore, the {higher frequency mode of branch (A) is identified as a bulk or Kittel mode while the lower non-harmonic frequencies are identified as predominantly edge modes. In the case of the branch (C), all modes are found to have mixed or hybrid bulk and edge characteristics.} 

{The calculated edge modes below the Kittel mode are not detectable experimentally since their absorption is substantially weaker than that of the uniform mode. The experimental observation of edge localized modes is highly dependent on both the feature shape and edges as well as on pinning effects \cite{Joe_APL13,Neusser_APL08,Neusser_PRB08,Ding_APL_2012}. The experimentally measured faint high frequency mode above the bulk mode [Fig.~\ref{Fig3}(a)] cannot be found in the results of the semianalytical model [Fig.~\ref{Fig3}(b)]. However, as we will see below, it also appears in the micromagnetic simulations suggesting that more complicated dynamics have to be taken into account.} 

Next we will focus on the low field regime where a hysteresis in the dynamic response 
is observed. Before the spectra were recorded, the following routine was used: First, the sample is saturated at a bias magnetic field of $H = -3100$~Oe in $-x$-direction. Then the magnetic field is reversed to a specific initializing field $H_\mathrm{ini}$. The final step is a lowering of the magnetic field to $H=0$~Oe and recording spectra while increasing the field gradually from 0~Oe to 1000~Oe. The corresponding viewgraphs are shown in Fig.~\ref{Fig4}. As is apparent from Fig.~\ref{Fig4}~(a), the two strong modes labeled as (A) and (B) do not coexist at zero field. This result agrees with the spectrum shown in Fig.~\ref{Fig3}, where no specific magnetic field-history procedure was applied. As the initialization field $H_\mathrm{ini}$ is increased, however, mode (B) starts to appear at smaller applied fields $H$, see Fig.~\ref{Fig4}~(b-c). At $H_\mathrm{ini}= 330$~Oe, Fig.~\ref{Fig4}~(c), both modes (A) and (B) do coexist at $H=0$~Oe. Increasing $H_\mathrm{ini}$ even further leads to a disappearance of mode (A) at $H=0$~Oe and only mode (B) persists. This result clearly shows the existence of field-dependent hysteresis in the spin-ice system.


In order to understand this behavior better, we carried out micromagnetic simulations using the experimental sample dimensions and separations. Details about the simulations can be found in the SM.
The insets in Fig.~\ref{Fig4} show the micromagnetic static equilibrium magnetization at zero-field,
following the protocol described above.  
We can clearly identify differences in the static equilibrium magnetization at $H=0$~Oe which concur with the experimentally observed appearance and coexistence of modes (A) and (B): Below 320~Oe [Fig.~\ref{Fig4}~(a)] the magnetization of the vertically oriented islands is not perfectly aligned {to the easy axis}. A formation of vortices and $S$-shaped magnetization configurations is found, which precludes the use of the simplified semianalytical treatment outlined above in the low-field regime. The horizontal islands are aligned in $-x$-direction; $H_\mathrm{ini}$ was not high enough to reverse the magnetization in those elements. Above 320~Oe [Fig.~\ref{Fig4}~(c,d)], however, the situation changes drastically: The vertical elements are perfectly aligned {to the easy axis} and the horizontal islands have partially flipped in the $+x$-direction. This field range where a change in the static equilibrium magnetization at zero-field occurs, agrees well with the experimentally observed onset of hysteresis in the dynamic response of the spin ices and a change in the mode structure at zero-field.

\begin{figure}[t]
\includegraphics[width=1\columnwidth]{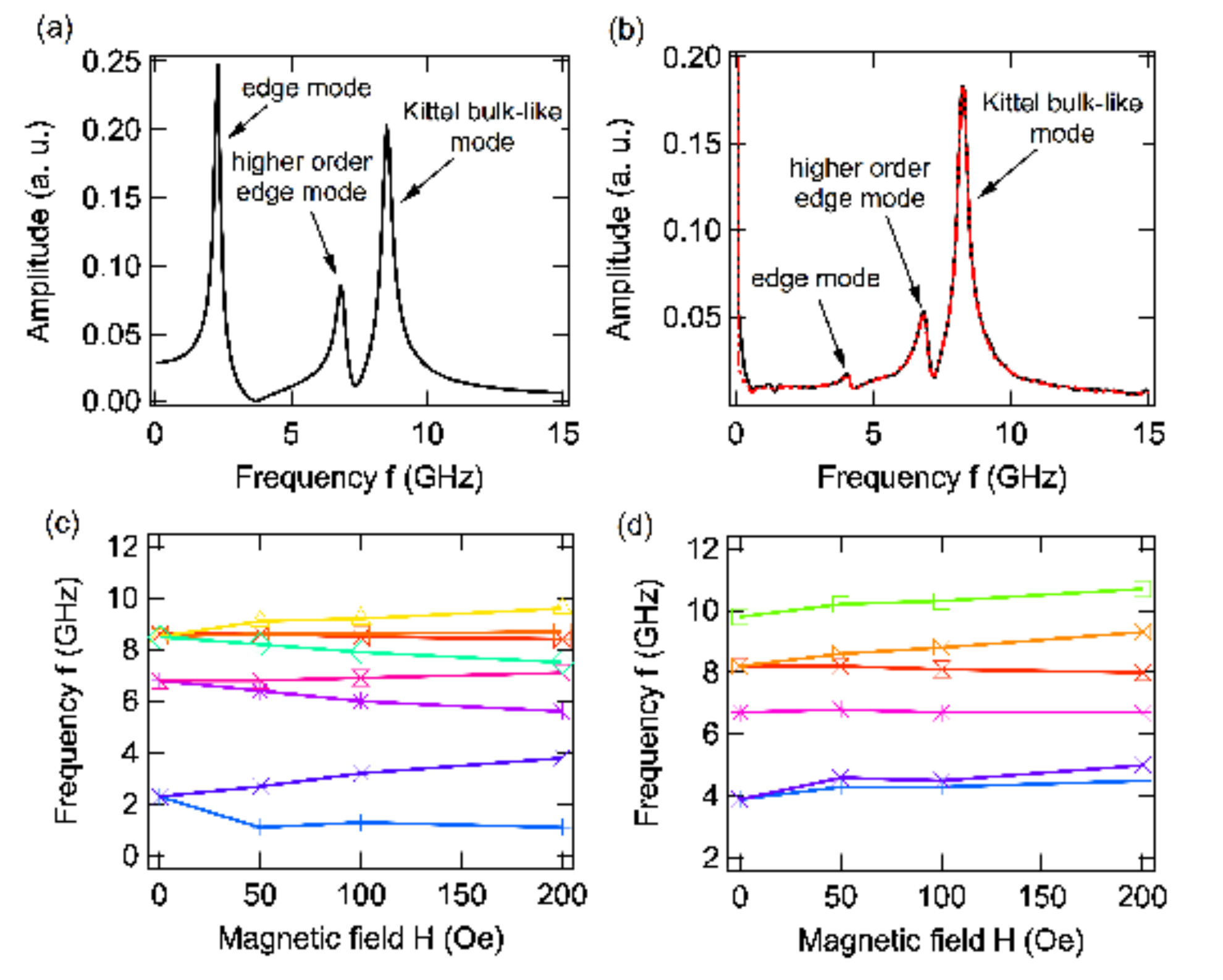}
\caption{\label{Fig5} (Color online) 
 Simulated magnetization dynamics. (a) Spectra of the 
type I state and (b) of the 
type II state. The horizontal islands are equivalent for the type I state at $H=0$~Oe and, thus, the spectrum is the same. This degeneracy can be lifted by applying a magnetic field in the horizontal direction. For the 
type II state, the magnetization components are different at $H=0$~Oe. The dashed-dotted line represents the vertical component of the magnetization.
(c,d) Simulated frequency-field relation of the different modes for (c) the 
ground state (type I) and for (d) the 
type II state.}
\end{figure}

Figure~\ref{Fig5} shows the simulated magnetization dynamics spectrum at $H=0$~Oe for (a) a 
type I spin ice, which is equivalent to the ground state of the system and (b) a 
type II spin ice. In general, three sets of even modes, two edge modes and one bulk-like (Kittel) mode can be observed. For the 
type I (type II) state, the mode at $f=2.3$~GHz ($3.9$~GHz) is the lowest edge mode, $f=6.9$~GHz ($6.7$~GHz) corresponds to a higher edge mode and $f=8.5$~GHz ($8.2$~GHz) is the Kittel mode, {in agreement} with the experiment [Fig.~\ref{Fig3}(a)]. Details about the time-dependent simulations are presented in the SM.

For the the 
type I state the horizontal islands are equivalent and, therefore, the spectrum is degenerate. This degeneracy can be lifted by applying a magnetic field in horizontal direction ($y$-direction). In this case, the modes split off and modes with a dynamical magnetization component in $x$-direction (vertically) in islands with a equilibrium magnetization in $-y$-direction will have a smaller resonance frequency with increasing magnetic field, see Fig.~\ref{Fig5}(c). On the other side, modes with an equilibrium magnetization direction in $+y$-direction show an increase of frequency with magnetic field. 
In contrast to the 
type I state, the type II state is non-degenerate since none of the islands are equivalent, see Fig.~\ref{Fig5}(b): The solid line represents islands with a dynamical component in $x$-direction and the dashed-dotted line is attributed to islands with a dynamical component in $y$-direction. As can be seen from Fig.~\ref{Fig5}(d), the frequency of all modes increases with increasing magnetic field. {Furthermore, we find higher frequency modes in the spectra for the 
type II state [Fig.~\ref{Fig5}(d)] corresponding to hybrid edge localized modes. This resembles the high frequency mode above the Kittel bulk-like mode found in experiment [Fig.~\ref{Fig3}(a)]. However, this mode is not visible in the minor loop experiment [Fig.~\ref{Fig4}] in agreement with the simulation results for the ground state, see Fig.~\ref{Fig5}(a).} 

In experiment (Fig.~\ref{Fig3} and Fig.~\ref{Fig4}), we only observe an increase of the strong high-frequency (Kittel) modes with the field. However, at low frequencies (edge modes), we find a {drop} of the frequency with the field. From the comparison with the micromagnetic simulations, we can conclude that both states (type I and type II) 
are present in the system and contribute to the dynamic response. As is apparent from Fig.~\ref{Fig5}(c), the splitting of the modes degenerate at zero field is largest for edge modes. This might be a reason why we experimentally only observe a splitting at low frequencies; for higher frequencies the absorption is much stronger and, thus, it might be that the splitting is not visible in experiment. We did not consider type III and IV spin-ice configurations in the simulations. However, our results indicate that it might eventually be possible to determine the spin-ice state by a resonance experiment.


In summary, we investigated magnetization dynamics of an artificial square spin-ice system in the GHz-regime using broadband ferromagnetic resonance spectroscopy. Complex spectra over the entire field range are found which are dictated by the magnetization states of individual islands, as is confirmed by a semianalytical model. An unusual hysteretic field behavior in the mode spectrum is found in the low-field regime which was not reported so far. Micromagnetic simulations reveal that this behavior originates from the underlying static equilibrium state of the system. The simulated dynamic magnetization spectra agree well with experiments and indicate that it {is} feasible to identify the spin-ice state by simple resonance measurements. In a similar manner, it might be possible to determine the magnetization state in crystalline spin ices. Furthermore, this work {opens up} routes for exploring novel phenomena and effects in artificial geometrically frustrated magnetic systems in the microwave frequency regime.

\begin{acknowledgments}

We thank Vitali Metlushko and Ralu Divan for assistance with the electron beam lithography. 
The work at Argonne was supported by the U.S. Department of Energy, Office of Science, Materials Science and Engineering Division. Lithography was carried out at the Center for Nanoscale Materials, an Office of Science user facility, which is supported by DOE, Office of Science, Basic Energy Science under Contract No. DE-AC02-06CH11357. E. Iacocca acknowledges support from the Swedish Research Council, Reg. No. 637-2014-6863. {We gratefully acknowledge the computing resources provided on Blues, a high-performance computing cluster operated by the Laboratory Computing Resource Center at Argonne National Laboratory.}

\end{acknowledgments}

\end{document}